\documentclass[%
twocolumn,
 floatfix,
 superscriptaddress,
nofootinbib,
 amsmath,amssymb,
 aps,
]{revtex4-2}

\usepackage{booktabs}
\usepackage[x11names,dvipsnames,svgnames]{xcolor}
\usepackage{graphicx}
\usepackage{dcolumn}
\usepackage{bm}
\usepackage{hyperref}
\hypersetup{
    colorlinks=true,
    linkcolor=red,
    citecolor=blue,
    filecolor=magenta,
    urlcolor=blue
}
\usepackage{ragged2e}
\usepackage{cleveref}
\usepackage[mathlines]{lineno}

\usepackage[normalem]{ulem}
\usepackage[caption=false,position=top]{subfig}
\usepackage[x11names]{xcolor}
\usepackage{custom}

\newcommand{\coef}{\alpha}
\newcommand{\sNN}{\sqrt{s_{\rm NN}}}

\allowdisplaybreaks
\begin{document}

\title{QCD critical surface from constant entropy contours}

\author{Hitansh Shah}
\affiliation{
 Department of Physics, University of Houston, Houston, TX 77204, USA
}

\author{Tristan Gyure}
\affiliation{
 Department of Physics, University of Houston, Houston, TX 77204, USA
}

\author{Anabella Leon}
\affiliation{
 Department of Physics, University of Houston, Houston, TX 77204, USA
}

\author{Francesco Di Clemente}
\affiliation{
 Department of Physics, University of Houston, Houston, TX 77204, USA
}

\author{Mauricio Hippert}
\affiliation{Centro Brasileiro de Pesquisas F\'isicas, Rua Dr. Xavier Sigaud 150, Rio de Janeiro, RJ, 22290-180, Brazil}

\author{Claudia Ratti}
\affiliation{
 Department of Physics, University of Houston, Houston, TX 77204, USA
}

\author{Volodymyr Vovchenko}
\affiliation{
 Department of Physics, University of Houston, Houston, TX 77204, USA
}

\date{\today}%

\begin{abstract}
We provide the first mapping of the critical surface in (2+1)-flavor QCD in the full $(T,\mu_B,\mu_Q,\mu_S)$ space, anchored on lattice QCD results at vanishing chemical potentials and obtained within an expansion along contours of constant entropy density.
In the pure $\mu_B$ direction, this framework yields a critical point at $(T_c,\mu_{B,c}) \simeq (114,\, 602)$ MeV.
Here we extend the construction to arbitrary directions in the three-dimensional chemical-potential space, parametrized by spherical coordinates $(\mu,\theta,\varphi)$, with the radial expansion truncated at $\mathcal{O}(\mu^2)$.
The resulting two-dimensional surface carries a direction-dependent critical temperature $T_c(\theta,\varphi)$ and baryochemical potential $\mu_{B,c}(\theta,\varphi)$, which quantify the shift of the critical point relative to the pure $\mu_B$ direction.
We find that $\mu_{B,c}$ increases by 40-100 MeV along the approximately strangeness neutral direction [$\mu_S \approx (0.15$--$0.33)\, \mu_B$, $\mu_Q \approx 0$] relevant for heavy-ion collisions, while the critical temperature stays essentially unchanged.
In the charge-neutral, weak-equilibrium direction~[$\mu_Q \approx -(0.05$--$0.1) \,\mu_B$, $\mu_S = 0$] relevant for neutron star mergers,
the critical point, and the associated first-order phase transition, remain present at essentially the same location in the $(T,\mu_B)$ plane.
We find no evidence for a critical point at large isospin densities, $|\mu_Q| / \mu_B \gtrsim 1$, relevant for cosmic trajectories in the early Universe, nor along the pure electric-charge or strangeness directions, at least outside the regions where pion or kaon condensation may occur.

\end{abstract}

\maketitle
\section{Introduction}
It is well-known from first-principles lattice QCD calculations that the transition from hadronic matter to the quark-gluon plasma (QGP) at baryonic chemical potential $\mu_B=0$ is a smooth analytic crossover at a pseudo-critical temperature $T_{\rm pc} = 155$--$158$ MeV \cite{Aoki:2006br,Borsanyi:2020fev,HotQCD:2018pds}. 
Whether this crossover ends at a critical point (CP) at high density, and if so, where, is an open question.
Direct lattice simulations cannot address it due to the fermion sign problem, which prevents calculations at finite baryon chemical potential $\mu_B$. 
Predictions for CP location, therefore, rely on alternative approaches.
These include functional QCD~\cite{Fu:2019hdw,Gao:2020fbl,Gunkel:2021oya}, holographic models \cite{Critelli:2017oub,Hippert:2023bel}, and lattice-based extrapolations \cite{Basar:2023nkp,Clarke:2024ugt,Shah:2024img}.
Several of these approaches predict a CP around $(T_c, \mu_{B,c}) \sim (110,600)$~MeV.
The predictions, if accurate, would place the CP in the vicinity of chemical freeze-out in heavy-ion collisions at moderate energies, $\sNN \simeq 3-5$~GeV~\cite{Andronic:2017pug,Lysenko:2024hqp}.

On the experimental side, the search for the CP drove the Beam Energy Scan program at RHIC~\cite{Bzdak:2019pkr,Du:2024wjm} and will continue at FAIR~\cite{CBM:2016kpk}.
Higher-order cumulants of the net-proton distribution have been the primary observable in the CP search, as they are highly sensitive to a nearby CP in equilibrium \cite{Stephanov:1999zu,Stephanov:2008qz}. 
Measurements by the STAR Collaboration~\cite{STAR:2020tga,STAR:2021iop,STAR:2025zdq} largely agree with non-critical baselines driven by baryon conservation and repulsive interactions~\cite{Vovchenko:2021kxx} at $\sNN \gtrsim 15-20$~GeV, largely ruling out the existence of the CP in the collider regime at RHIC-BES.
Clear deviations from the baselines emerge at $\sNN \lesssim 15$~GeV, but decisive conclusions are elusive due to the many challenges associated with interpreting fluctuation measurements, especially as the collision energy decreases (see Ref.~\cite{Koch:2025cog} for a recent review).
Interestingly, deviations in the same energy range also emerge in other potentially relevant observables, such as mean $p_T$ fluctuations~\cite{STAR:2026vjv}. 

\begin{figure*}[t]
    \includegraphics[width=\linewidth]{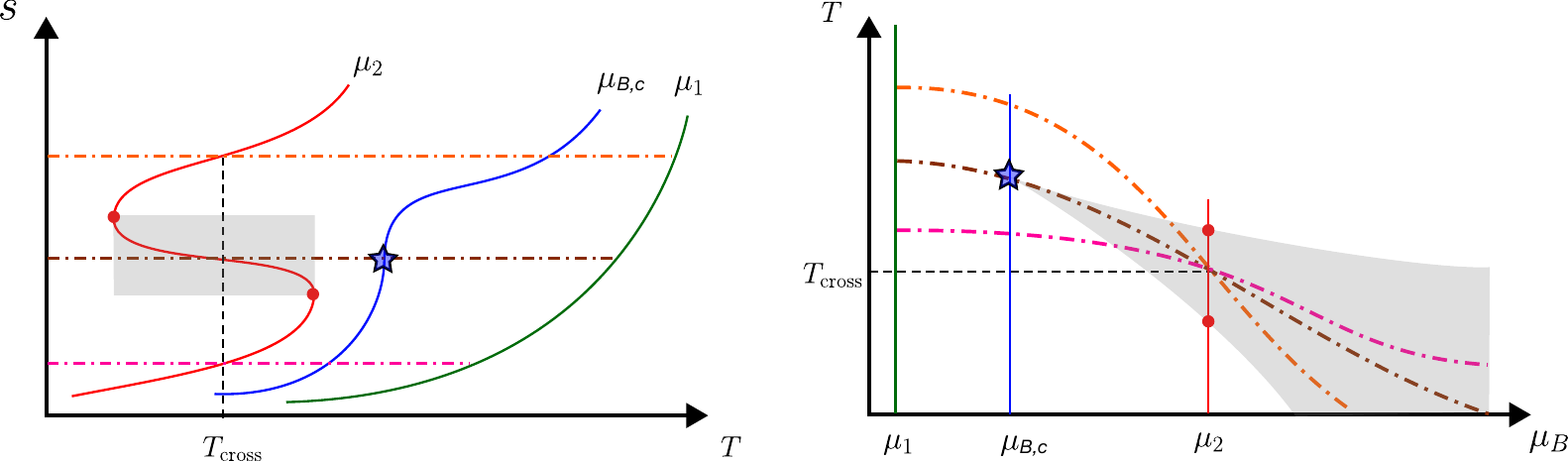}
   \caption{ \justifying \small
\textit{Left panel:} Entropy described as a function of the temperature for three
representative baryon chemical potentials, with
$\mu_1 < \mu_{B,c} < \mu_2$.
\textit{Right panel:} The corresponding constant-entropy trajectories mapped
in the $(T, \mu_B)$ plane.
The blue star demonstrates the location of the critical point, while the shaded
region indicates the spinodal domain, and the red dots highlight the
spinodal points at $\mu_B = \mu_2$.  Figure is acquired from
\cite{Shah:2024img}.
}
    \label{fig:cartoon_firstorder}
\end{figure*}
In Ref. \cite{Shah:2024img}, we proposed locating the CP by tracking contours of constant entropy density in the $(T,\mu_B)$ plane, extrapolating from $\mu_B = 0$ via a Taylor series.

Truncated at $\mathcal{O}(\mu_B^2)$, the method gives $(T_c, \mu_{B,c}) = (114 \pm 7,\, 602 \pm 62)$ MeV.
The Wuppertal-Budapest collaboration later applied the method under strangeness-neutral conditions to exclude a CP below $\mu_B = 450$ MeV \cite{Borsanyi:2025dyp}.
In Ref.~\cite{Shah:2026pue}, the method was expanded to reconstruct other thermodynamic quantities via the integration of entropy density, giving the access to the full EoS.
The method was also tested against solvable effective QCD theories in Refs.~\cite{Marczenko:2025znt,Shah:2026pue}, indicating that it accurately reproduces the CP at $\mu_B \simeq 600$~MeV predicted by functional methods~\cite{Gao:2020fbl,Lu:2025cls} and holography~\cite{Critelli:2017oub,Hippert:2023bel}, but can also yield a spurious CP at higher $\mu_B$ and lower $T$~\cite{Marczenko:2025znt}.

In this work, we extend the method to the four-dimensional $(T,\mu_B,\mu_Q,\mu_S)$ space, where $\mu_Q$ and $\mu_S$ denote the electric-charge and strangeness chemical potentials. 
To this end, we use spherical coordinates $(\mu,\theta,\varphi)$ in the chemical potential space, originally introduced in Ref.~\cite{Abuali:2025tbd} within a $T'$-expansion, and search for entropy crossings indicating the CP across all angular directions.
The resulting critical points trace a two-dimensional surface with a direction-dependent critical temperature $T_c(\theta,\varphi)$ and baryochemical potential $\mu_{B,c}(\theta,\varphi)$.
The surface is approximately elliptic in the $(\mu_B,\mu_S)$ plane and hyperbolic in the $(\mu_B,\mu_Q)$ plane, controlled by the signs and magnitudes of the off-diagonal susceptibilities $\chi_{11}^{BS}$, $\chi_{11}^{BQ}$ and the diagonal $\chi_2^Q$, $\chi_2^S$.
In particular, we determine the existence and location of the CP  along several physically relevant directions in chemical-potential space. These include strangeness-neutral matter~($\mu_S/\mu_B > 0$, $\mu_Q \approx 0$), relevant for  heavy-ion collisions; charge-neutral matter in weak equilibrium~($\mu_Q/\mu_S < 0$, $\mu_S = 0$),  relevant for neutron stars; and matter with large lepton flavor asymmetry and isospin density~($|\mu_Q| / \mu_B \gtrsim 1$, $\mu_S = 0$), relevant for the early Universe.
Regarding the $\mu_S$ and $\mu_Q$ directions, we expect the validity of the construction to be restricted to the region where Bose--Einstein condensation of pions and kaons is absent, i.e. $|\mu_Q| \lesssim m_\pi$, $|\mu_S| \lesssim m_K$, and $|\mu_Q + \mu_S| \lesssim m_K$.

To our knowledge, this is the first such mapping of the QCD critical surface from lattice inputs at vanishing chemical potentials.
Although four-dimensional lattice-based state equations have been developed previously~\cite{Noronha-Hostler:2019ayj,Monnai:2024pvy,Abuali:2025tbd}, they were based on explicit Taylor expansions for pressure (or shifted temperature $T'$ in \cite{Abuali:2025tbd}) and could not incorporate a CP by construction.
Functional methods~\cite{Fischer:2026uni} are in principle capable of such a calculation, but so far this has been explored only for specific physically relevant directions, such as strangeness neutrality~\cite{Fu:2026qnl}.

The manuscript is organized as follows.
Section \ref{sec:method} reviews the constant entropy density contour method and presents its extension to the three-dimensional chemical potential space $(\mu_B,\mu_Q,\mu_S)$.

Section \ref{sec:results} describes the lattice QCD input and the propagation of its uncertainties, and the resulting critical surface.
Discussion and conclusions in Sec.~\ref{sec:Conclusions} close the article.

\section{Methodology}
\label{sec:method}
\subsection{Contours of constant entropy density} \label{sec:s_contours_method}

The entropy-density contour method, introduced in Ref.~\cite{Shah:2024img}, locates the QCD critical point by tracking lines of constant entropy density in the $(T,\mu_B)$ plane.
The motivation is the behavior of $s$ near a first-order phase transition: 
at the mean-field level, $s$ becomes a multi-valued function of $T$ and $\mu_B$ in the thermodynamic limit, describing stable, metastable, and unstable (spinodal) branches.

This is illustrated in Fig. \ref{fig:cartoon_firstorder}.
\begin{figure*}
    \centering
    \includegraphics[width=\textwidth]{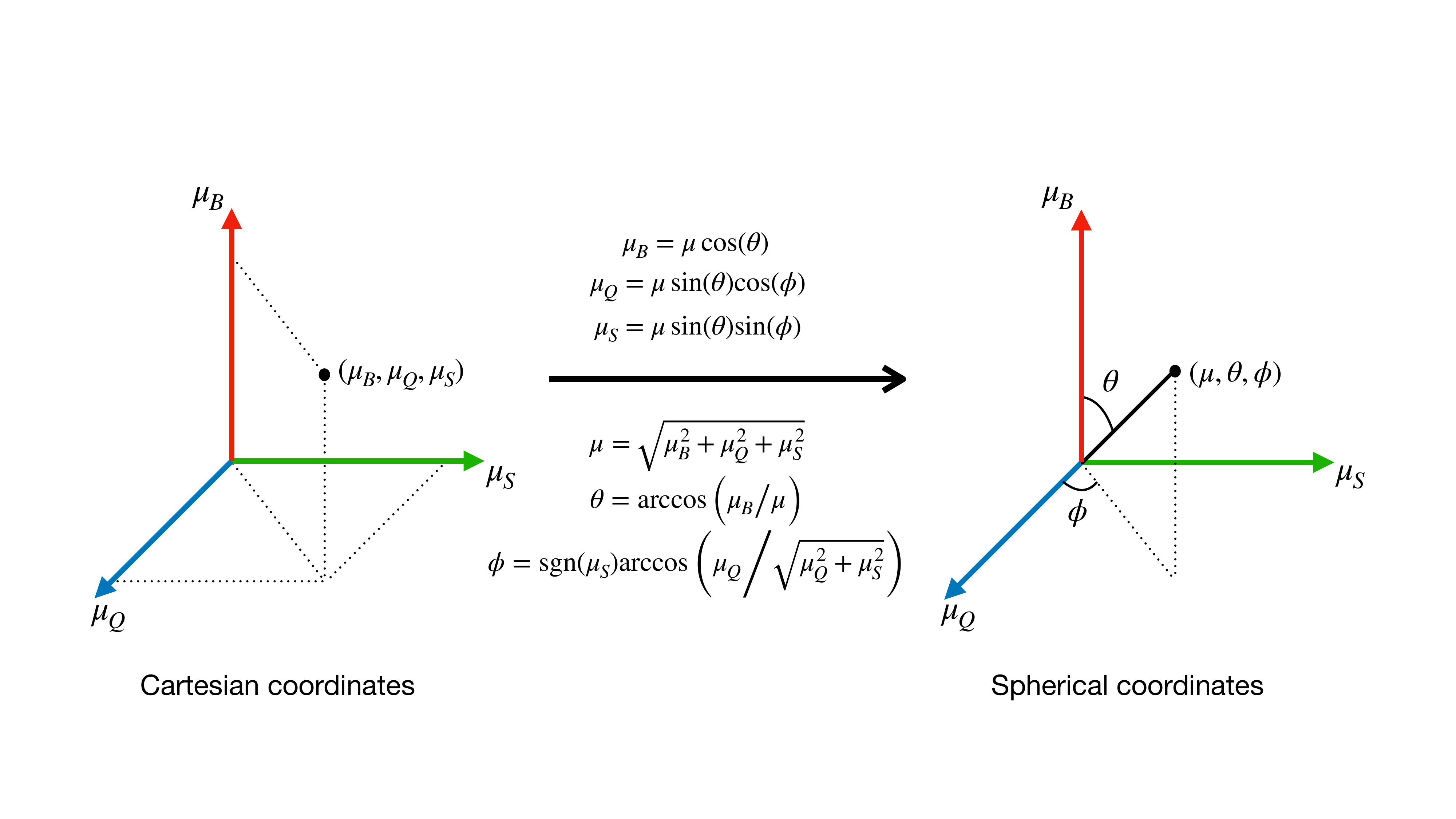}
    \vspace{-1.5cm}
    \caption{ The three-dimensional space of conserved charge chemical
    potentials in Cartesian $(\mu_B,\mu_Q,\mu_S)$ (left) and
    spherical $(\mu,\theta,\varphi)$ (right) coordinates. The
    spherical parameterization reduces the four-dimensional problem
    $(T,\mu_B,\mu_Q,\mu_S)$ to a family of two-dimensional problems
    $(T,\mu)$ indexed by the angles $(\theta,\varphi)$, enabling a
    one-dimensional extrapolation in any direction of chemical
    potential space.
    }
    \label{fig:mapping}
\end{figure*}
The left panel of Fig.~\ref{fig:cartoon_firstorder} displays the entropy
density $s$ as a function of temperature $T$ at three
values of the baryon chemical potential. At low $\mu_B = \mu_1$, the entropy density
increases monotonically with $T$, suggesting a smooth crossover. At the
critical chemical potential $\mu_{B,c}$, the slope
$(\partial s / \partial T)_{\mu_B}$ diverges at the critical temperature $T_c$,
indicating the point of the second-order transition. For large baryon chemical potential
$\mu_B = \mu_2 > \mu_{B,c}$, a single $(T,\mu_B)$ pair admits three
distinct values of $s$, and the projection of the constant-entropy
contours onto the $(T,\mu_B)$ plane yields intersecting trajectories
spanning the full spinodal region of the first-order phase transition,
shown in gray in the right panel.

Since direct lattice QCD simulations are restricted to vanishing
chemical potential, the constant entropy density contours must be
accessed through a Taylor expansion in $\mu_B$ anchored at $\mu_B=0$:

\begin{equation}
    T_s(\mu_B; T_0) \approx T_0 + \sum_{n=1}^N \coef_{2n}(T_0)
    \,\frac{\mu_B^{2n}}{(2\,n)!}
    + \mathcal{O}\left(\mu_B^{2(N+1)}\right),
    \label{eq:expSsup}
\end{equation}
where the expansion coefficients are evaluated along contours of
fixed $s$,
\begin{equation}
\coef_{2n}(T_0) = \left. \left(\frac{\partial^{2n} T}{\partial
\mu_B^{2n}}\right)_s \right|_{T = T_0, \mu_B = 0}.
\end{equation}

Because the net baryon density vanishes identically at $\mu_B = 0$,
charge-conjugation symmetry enforces $\rho_B = 0$ at all temperatures
on this axis, causing all odd-order coefficients to vanish. The
leading non-trivial coefficient is therefore:

\begin{equation}
\label{eq:app:C2}
\coef_2(T_0) = \left. \left(\frac{\partial^2 T}{\partial
\mu_B^2}\right)_s \right|_{T=T_0, \mu_B = 0}
\!\!\!\!\!\!= -\frac{2T_0 \chi_2^{B}(T_0) + T_0^2 \chi_2^{B'}(T_0)}{s'(T_0)},
\end{equation}
and the expansion truncated at $\mathcal{O}(\mu_B^2)$ reads
\begin{equation}
\label{eq:TsO2}
T_s(\mu_B;T_0) = T_0 + \coef_{2}(T_0) \frac{\mu_B^2}{2}.
\end{equation}

The critical point is identified via two conditions that follow from
its character as an inflection point of the equation of state: the
entropy slope diverges, $(\partial T/\partial s)_{\mu_B} = 0$, and
the curvature vanishes, $(\partial^2 T/\partial s^2)_{\mu_B} = 0$.
Translating these into conditions on the expansion, and noting that
$(\partial s/\partial T_0)_{\mu_B=0}$ is strictly positive at all
finite temperatures, yields
$\left(\partial T_s/\partial T_0\right)_{\mu_B} = 0$ and
$\left(\partial^2 T_s/\partial T_0^2\right)_{\mu_B} = 0$.
Denoting by $\mu_{B,c}$ and $T_{0,c}$ the values at which these
conditions are simultaneously satisfied, the first equation yields
\begin{equation}
\label{eq:spinodal}
1 + \coef_2'(T_{0,c}) \frac{\mu_{B,c}^2}{2} = 0
\quad \Rightarrow \quad
\mu_{B,c} = \sqrt{-\frac{2}{\coef_2'(T_{0,c})}},
\end{equation}
which also governs the spinodal boundaries at $\mu_B > \mu_{B,c}$,
where it admits two solutions in $T_0$. The second condition requires
\begin{equation}
\label{eq:T0c}
\coef_2''(T_{0,c}) = 0,
\end{equation}
which uniquely determines $T_{0,c}$. The full critical point location
is then obtained by solving Eq.~\eqref{eq:T0c} for $T_{0,c}$,
substituting into Eq.~\eqref{eq:spinodal} to find $\mu_{B,c}$, and
evaluating $T_c = T_s(\mu_{B,c};T_{0,c})$ via Eq.~\eqref{eq:TsO2}.
For $\mu_B > \mu_{B,c}$, the Maxwell equal-area construction on the
entropy density yields the phase coexistence curve.

\subsection{Extension to four dimensions}

The full thermodynamic space of (2+1)-flavor QCD is spanned by the temperature and
the chemical potentials for the three conserved charges $(T,\mu_B,\mu_Q,\mu_S)$.
To apply the entropy density contour expansion across this space, we
span the three-dimensional chemical potential subspace using
spherical coordinates $(\mu,\theta,\varphi)$~(Fig.~\ref{fig:mapping}):
\begin{align}
\label{eq:mu_conversions}
    \mu_B &= \mu \cos\theta, \\
    \mu_Q &= \mu \sin\theta\cos\varphi, \\
    \mu_S &= \mu \sin\theta\sin\varphi.
\end{align}

The reparameterization reduces the four-dimensional problem to a family of two-dimensional problems $(T,\mu)_{\theta,\varphi}$, one for
each fixed direction $(\theta,\varphi)$ in chemical potential space.

The spherical coordinate parametrization was previously employed in Ref.~\cite{Abuali:2025tbd} to construct the QCD equation of state via the $T'$ expansion scheme. 
There are two key differences between the current scheme and the $T'$ expansion scheme.
First, motivated by their quadratic structure at imaginary chemical potentials in lattice QCD~\cite{Borsanyi:2025dyp}, we expand the contours of constant entropy density $s$ instead of the scaled baryon density $\rho_B / (\mu_B/T)$ in \cite{Abuali:2025tbd}. 
Second, we formulate the expansion in an \emph{implicit} form: we fix $T_0$ and $\mu_B$ and calculate the resulting temperature $T_s$ at a finite $\mu_B$ via Eq.~\eqref{eq:expSsup}.
Instead, in the $T'$-expansion scheme one fixes $(T,\mu_B)$ directly and then computes $T_0$.
It is the fact that the temperature is implicit in our scheme that allows us to obtain a multi-valued behavior of $s$ and describe a first-order phase transition with a CP.

Writing the pressure as
$P(T,\mu,\theta,\varphi)$, the generalized susceptibilities along a
given direction are defined as
\begin{equation}
    X_n^{\theta,\varphi} (T) = \frac{\partial^n (p/T^4)}{\partial
    (\mu/T)^n} \Bigg|_{\theta,\varphi}.
\end{equation}
The first- and second-order generalized susceptibilities decompose into a linear combination of the standard lattice QCD susceptibilities,
\begin{align}
\label{eq:X1}
    X_1^{\theta,\varphi}(T) & =
    c_{\theta} \chi_1^B(T) + s_{\theta} c_{\varphi} \chi_1^Q(T)
    + s_{\theta} s_{\varphi} \chi_1^S(T) \, , \\[1em]
\label{eq:X2}
    X_2^{\theta,\varphi}(T) & =
    c_{\theta}^2\chi_2^B(T)
    + s_{\theta}^2c_{\varphi}^2\chi_2^Q(T)
    + s_{\theta}^2 s_{\varphi}^2\chi_2^S(T) \nonumber \\
    & \quad + 2c_{\theta}s_{\theta}c_{\varphi}\chi_{11}^{BQ}(T)
    + 2c_{\theta}s_{\theta}s_{\varphi}\chi_{11}^{BS}(T)  \nonumber \\
    & \quad 
    + 2s_{\theta}^2c_{\varphi}s_{\varphi}\chi_{11}^{QS}(T)\, ,
\end{align}
where $s_{\gamma} \equiv \sin\gamma$, $c_{\gamma} \equiv \cos\gamma$,
and
\begin{equation}
\label{eq:XBSQ}
\chi_{lmn}^{BQS}~=~\frac{\partial^{l+m+n} P(T,\mu_B,\mu_Q,\mu_S)/T^4}{\partial(\mu_B/T)^l \,\partial(\mu_Q/T)^m \,\partial(\mu_S/T)^n}~\,.
\end{equation}

Note that $X_1^{\theta,\varphi}$ vanishes at $\mu = 0$ because all
individual charge densities are zero on the
$\mu_B = \mu_Q = \mu_S = 0$ axis.

The entropy density contour expansion in the generalized radial
direction $\mu$ takes the same form as in the one-dimensional case:
\begin{equation}
\label{eq:Ts02_4D}
    T_s^{\theta,\varphi}(T_0, \mu) = T_0
    + \frac{\mu^2}{2}\, \alpha_2^{\theta,\varphi}(T_0),
\end{equation}
where the direction-dependent expansion coefficient is
\begin{align}
\label{eq:alpha2_4D}
    \alpha_2 (T_0;\theta,\varphi) & =
    \left. \left(\frac{d^2 T}{d\mu^2}\right)_{\theta,\varphi} \right|_{\mu=0} \nonumber \\ &
    = -\frac{\partial_{T_0} [T_0^2 X_2^{\theta,\varphi}(T_0)]}{s'(T_0)} \nonumber \\ &
    = -\frac{2 T_0 X_2^{\theta,\varphi}(T_0)
    + T_0^2 \partial_{T_0} X_2^{\theta,\varphi}(T_0)}{s'(T_0)},
\end{align}
with $X_2^{\theta,\varphi}$ defined in Eq.~\eqref{eq:X2}. The
critical point conditions are carried over from the one-dimensional case, the only modification being the angular dependence of $\alpha_2$.
For each direction $(\theta,\varphi)$, the critical point is located
by solving
\begin{equation}
    \label{eq:CP_4D}
    \mu_c^{\theta,\varphi} = \sqrt{-\frac{2}{\alpha_2'(T_{0,c}^{\theta,\varphi};\theta,\varphi)}},
    \qquad \alpha_2''(T_{0,c}^{\theta,\varphi};\theta,\varphi) = 0,
\end{equation}
where primes denote derivatives with respect to $T_0$. Once $\mu_c^{\theta,\varphi}$
and $T_{0,c}^{\theta,\varphi}$ are determined, the critical temperature follows from
$T_c^{\theta,\varphi} = T_s(T_{0,c}^{\theta,\varphi},\mu_c^{\theta,\varphi};\theta,\varphi)$ via
Eq.~\eqref{eq:Ts02_4D}. Following Eq. \eqref{eq:mu_conversions}, one can calculate the critical conserved charge chemical potential $\mu_{B_c}^{\theta,\varphi},\mu_{S_c}^{\theta,\varphi},\mu_{Q_c}^{\theta,\varphi}$ using the generalized critical chemical potential $\mu_c^{\theta,\varphi}$ for the corresponding direction.

Before proceeding to the calculations, let us first discuss caveats of the expansion scheme. 
The expansion is truncated at the 2nd order, and we do not take into account the truncation error of the expansion in this work. 
This is challenging as it would require higher-order conserved charge susceptibilities up to 4th order and their temperature derivatives, which are not yet available from the lattice with sufficient precision.
In the pure $\mu_B$ direction, the scheme predicts a CP $(T_c, \mu_{B,c}) = (114,\, 602)$ MeV at order $\mathcal{O}(\mu_B^2)$~\cite{Shah:2024img}.
This prediction is consistent with those of several other approaches, such as functional QCD and holography, but it does not prove definitively that the CP exists there.
Instead, the multi-dimensional construction presented here primarily shows how the CP would move in the $\mu_S$ and $\mu_Q$ directions if the CP exists in the pure $\mu_B$ direction in QCD and is located where predicted by the method and the aforementioned approaches, such as functional QCD and holography.

Another caveat relates to the known expected non-analyticity at large electric and strangeness charge chemical potentials due to Bose--Einstein condensation.

Pion condensation at finite $|\mu_Q| \gtrsim m_\pi$ and $T \lesssim 160$ MeV is expected and confirmed by lattice QCD simulations at finite isospin density~\cite{Brandt:2017oyy}. 
Similarly, kaon condensation is expected for $|\mu_S| \gtrsim m_K$ or $|\mu_S + \mu_Q| \gtrsim m_K$.

The expansion in its current form does not describe Bose-Einstein condensation, so one should treat the results with care when going to chemical potentials larger than the masses of these bosons. 

\section{Results}
\label{sec:results}

\subsection{Lattice input}

To obtain the expansion coefficient $\alpha_2^{\theta,\varphi}(T_0)$ in the three-dimensional space of chemical potentials $\mu_B$, $\mu_Q$, and $\mu_S$, we use the continuum-estimated conserved charge susceptibilities from lattice QCD presented in Ref.~\cite{Abuali:2025tbd} as input into Eq.~\eqref{eq:alpha2_4D}. 
As the framework requires higher-order temperature derivatives, we parametrize four second-order susceptibilities which are $\chi_2^B(T_0),\chi_2^S(T_0),\chi_2^Q(T_0)$ and $\chi_{11}^{QS}(T_0)$ along with
the entropy density $s(T_0)$ at $\mu = 0$. For the remaining two susceptibilities, we use the isospin symmetry condition which is imposed on the lattice QCD results at zero $\mu$, through which we acquire the relations:
\begin{align}
2\chi_{11}^{BQ}(T_0) - \chi_2^B(T_0) - \chi_{11}^{BS}(T_0) &= 0, \\
2\chi_{11}^{QS}(T_0) - \chi_2^S(T_0) - \chi_{11}^{BS}(T_0) &= 0.
\end{align}
Through these relations, the $\chi_{11}^{BQ}(T_0)$ and $\chi_{11}^{BS}(T_0)$ results as functions of the temperature are obtained. 
We preserve the same parametrization for $s(T_0)$ and $\chi_2^B(T_0)$ as in Ref.~\cite{Shah:2024img} to stay consistent with the existing pure $\mu_B$ results.
For the other three second-order susceptibilities, we use the same functional form as for $\chi_2^B$ but refit the parameters to match the lattice data.
Section~\ref{apdx:supplemental} in the Appendix provides the details of the parametrization and the resulting parameter values and their covariances.

\subsection{Critical line in the $\mu_B$-$\mu_S$ plane}

\subsubsection{$\mu_S/\mu_B$ scan}

\begin{figure*}[t]
    \centering
    \includegraphics[width=0.4\textwidth]{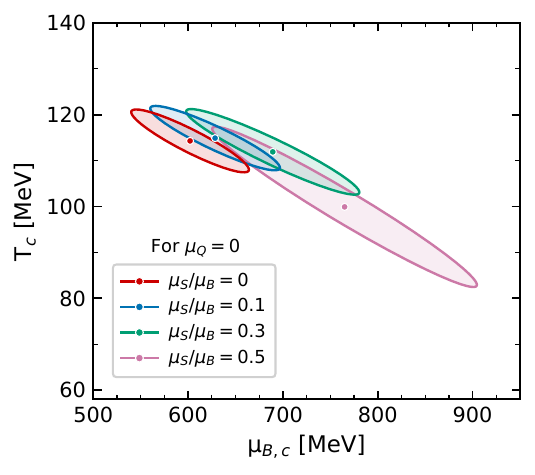}
    \hfill
    \includegraphics[width=0.56\textwidth]{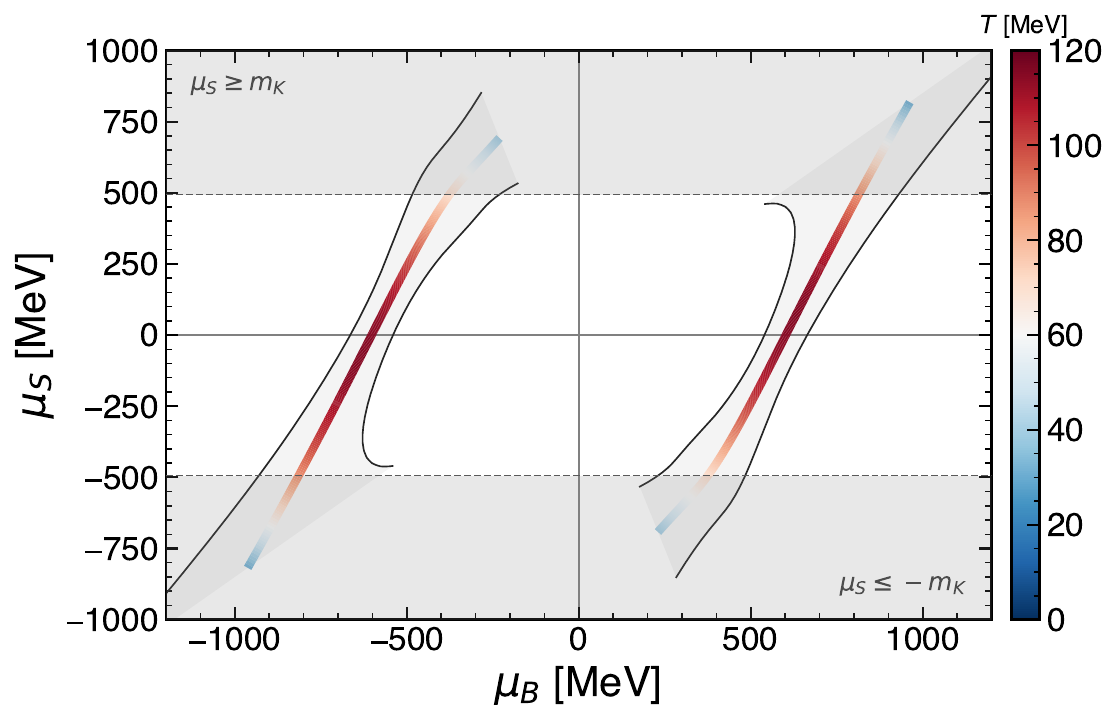}
    \caption{
    Critical point structure in the $(\mu_B,\mu_S)$ plane at $\mu_Q = 0$.
    The left panel shows the critical point ellipses in the $(T,\mu_B)$-plane with $1\sigma$ error bar for different values of $\mu_S/\mu_B$.
    The right panel depicts the critical line in the $(\mu_B, \mu_S)$-plane, where color indicates the critical temperature, and solid black lines indicate the uncertainty band.
    The gray shaded region corresponds to $|\mu_S|>m_K$, where kaon condensation is expected and where the present expansion should be treated with caution. 
    }
    \label{fig:muBmuS}
\end{figure*}

Now that we have defined the susceptibilities and entropy density, we can calculate the expansion coefficient
$\alpha_2(T, \theta, \varphi)$ using Eq.~\eqref{eq:alpha2_4D} for each direction in the 4D space, calculate its derivatives, and solve equations~\eqref{eq:CP_4D} to obtain the CP location. 
Figure~\ref{fig:muBmuS} shows the movement of the critical point in the $\mu_B$-$\mu_S$ plane at $\mu_Q = 0$, which is obtained by fixing $\varphi = 90^\circ$.
The ellipses in the left panel represent a $1\sigma$ uncertainty on the CP location for fixed values of the $\mu_S/\mu_B$ ratio, 
reflecting the linear propagation of the uncertainties in the lattice QCD input through automatic differentiation.
The right panel depicts the critical line, with the color indicating the critical temperature, indicating an elliptical structure in the $(\mu_B,\mu_S)$ plane.
The charge-conjugation symmetry of the QCD partition function is built into our expansion through the spherical coordinates by construction, and this is reflected in the right panel of Fig.~\ref{fig:muBmuS} by the invariance of the results with respect to the $(\mu_B,\mu_S) \to (-\mu_B, -\mu_S)$ transformation.

We observe that the critical temperature is nearly constant at small values of the $\mu_S / \mu_B$ ratio, and then decreases as $|\mu_S| / \mu_B$ increases.
The shift in $\mu_{B,c}$ is approximately linear in $\mu_S / \mu_B$ at moderate values of the ratio, $|\mu_S| / \mu_B \lesssim 0.5$.
The shifts at small $\mu_S / \mu_B$ are mainly driven by the baryon-strangeness correlator, $\chi_{BS}$.
The uncertainty in the critical point estimate increases significantly at higher $\mu_S/\mu_B$ values, while the temperature drops. 
At $\mu_S/\mu_B \gtrsim 1$, the extracted $T_c$ turns negative, and the equations~\eqref{eq:CP_4D} do not contain a solution at physical (positive) values of the temperature.
This indicates the disappearance of the CP with increasing $\mu_S/\mu_B$.
The uncertainties in the critical point location for different ratios of $\mu_S/\mu_B$ are provided in Table \ref{tab:muBmuS_CP}.

\begin{table*}[ht]
\centering
\begin{tabular}{|c|c|c|c|c|c|}
\hline
$\mu_S/\mu_B$ & $T_{0c}$ (MeV) & $T_c$ (MeV) & $\mu_{B,c}$ (MeV) & $\Delta T_c$ (MeV) & $\Delta \mu_{B,c}$ (MeV) \\
\hline
$-0.5$ & $139.5 \pm 2.6$ & $100.7 \pm 11.8$ & $496.5 \pm 59.2$  & $-13.6 \pm 6.3$ & $-105.6 \pm 29.4$ \\
$-0.4$ & $139.9 \pm 2.4$ & $104.3 \pm 10.2$ & $514.9 \pm 57.4$  & $-10.0 \pm 4.5$ & $-87.2 \pm 24.3$  \\
$-0.3$ & $140.2 \pm 2.2$ & $107.5 \pm 8.8$  & $534.5 \pm 56.3$  & $-6.8 \pm 2.9$  & $-67.6 \pm 18.8$  \\
$-0.2$ & $140.5 \pm 2.1$ & $110.4 \pm 7.8$  & $555.4 \pm 56.5$  & $-3.9 \pm 1.6$  & $-46.7 \pm 13.0$  \\
$-0.1$ & $140.7 \pm 2.0$ & $112.7 \pm 7.1$  & $577.8 \pm 58.2$  & $-1.6 \pm 0.7$  & $-24.2 \pm 6.8$   \\
$0$    & $140.9 \pm 1.9$ & $114.3 \pm 6.9$  & $602.1 \pm 62.0$  & $0.0 \pm 0.0$   & $0.0 \pm 0.0$     \\
$0.15$ & $141.2 \pm 1.8$ & $114.7 \pm 7.3$  & $642.6 \pm 72.6$  & $0.5 \pm 0.9$   & $40.5 \pm 12.7$    \\
$0.1$  & $141.1 \pm 1.9$ & $114.9 \pm 7.0$  & $628.4 \pm 68.4$  & $0.6 \pm 0.5$   & $26.4 \pm 8.0$    \\
$0.2$  & $141.3 \pm 1.8$ & $114.2 \pm 7.8$  & $657.3 \pm 77.8$  & $-0.1 \pm 1.3$  & $55.3 \pm 18.0$   \\
$0.3$  & $141.5 \pm 1.8$ & $111.9 \pm 9.3$  & $689.3 \pm 91.3$  & $-2.4 \pm 2.7$  & $87.2 \pm 31.7$   \\
$0.333$  & $141.5 \pm 1.8$ & $110.7 \pm 10.1$ & $700.7 \pm 97.0$  & $-3.6 \pm 3.5$  & $98.6 \pm 37.4$   \\
$0.4$  & $141.6 \pm 1.7$ & $107.4 \pm 12.2$ & $724.9 \pm 110.9$ & $-6.9 \pm 5.5$  & $122.8 \pm 51.3$  \\
$0.5$  & $141.7 \pm 1.7$ & $99.9 \pm 17.5$  & $765.0 \pm 139.5$ & $-14.4 \pm 10.7$ & $162.9 \pm 80.4$ \\
\hline
\end{tabular}
\caption{The location of the critical point up to $1\sigma$ for different ratios of $\mu_S/\mu_B$. Here, $\Delta T_c = T_c-T_c(\mu_S{=}0)$ and $\Delta\mu_{B,c} = \mu_{B,c}-\mu_{B,c}(\mu_S{=}0)$ are the shifts relative to the pure-$\mu_B$ critical point.}
\label{tab:muBmuS_CP}
\end{table*}

\subsubsection{Strangeness neutrality and heavy-ion collisions}

A non-zero $\mu_S$ allows one to incorporate the condition of strangeness neutrality, $n_S = 0$, on the net-strangeness density, which is a physically relevant condition for heavy-ion collisions.
We note that calculating a conserved-charge density requires the equation of state at finite $(\mu_B,\mu_Q,\mu_S)$, since it is given by the pressure derivative, i.e. $n_S = \partial P/\partial \mu_S$.
The method presented here provides the entropy density $s(T,\mu_B,\mu_Q,\mu_S)$, rather than the pressure $P(T,\mu_B,\mu_Q,\mu_S)$. Calculating the pressure and thus the full four-dimensional equation of state requires integrating the entropy density at fixed chemical potentials and fixing the integration constant, as was done in \cite{Shah:2026pue} for the pure $\mu_B$ direction.
While this procedure is beyond the scope of the present work, 
we estimate the relevant $\mu_S / \mu_B$ ratio for strangeness neutrality to estimate the location of the CP.

At high temperatures, where QCD thermodynamics is approximated by a quark gas, strangeness neutrality corresponds to setting the strange quark chemical potential to zero, $\mu_s = 0$, which corresponds to $\mu_S = \mu_B / 3$.
The case $\mu_s = 0$ is instructive also because it can be studied in various theoretical approaches, such as functional QCD~\cite{Gunkel:2021oya}, more straightforwardly than $n_S = 0$.
Under these conditions, the CP is located at $(T_c, \mu_{B,c}) = (111 \pm 10, 701 \pm 97)$~MeV, as seen in Table \ref{tab:muBmuS_CP}.
We note that, while this CP location appears to be consistent, within errors, with the one in the pure-$\mu_B$ direction, the errors are correlated because they are based on the same lattice QCD input.
Accounting for this correlation in the error propagation, we find that the shift in the temperature is $\Delta T_c = -3.6 \pm 3.5$~MeV, i.e. approximately a one-$\sigma$ effect.
However, the shift in $\mu_{B,c}$ is statistically significant, namely $\Delta \mu_{B,c} = 98.6 \pm 37.4$~MeV.

The value $\mu_S = \mu_B / 3$ corresponds to strangeness neutrality in the high-temperature limit.
At finite temperatures, this value is expected to be smaller, as indicated from lattice QCD at small baryon densities within $T'$-expansion scheme~\cite{Borsanyi:2022qlh}.
To estimate such a value of $\mu_S/\mu_B$ we perform an HRG model calculation at $(T,\mu_B) = (100, 600)$~MeV using \texttt{Thermal-FIST}~\cite{Vovchenko:2019pjl}, which yields $\mu_S / \mu_B \approx 0.15$-$0.20$ depending on the details of the HRG model.
Taking $\mu_S/\mu_B = 0.15$ as a lower estimate of the strangeness-neutral $\mu_{S,c}/\mu_{B,c}$ ratio, the CP location is $(T_c, \mu_{B,c}) = (115 \pm 7, 650 \pm 73)$~MeV, representing a $(\Delta T_c, \Delta \mu_{B,c}) = (0.5 \pm 0.8, 40 \pm 13)$~MeV shift of the CP location relative to the pure-$\mu_B$ case. 
We note that a strangeness neutral CP was recently analyzed within functional QCD~\cite{Fu:2026qnl}, where the corresponding shift relative to pure-$\mu_B$ direction can be inferred as $(\Delta T_c, \Delta \mu_{B,c}) = (-10, 52)$~MeV.
The upward shift $\mu_{B,c}$ is consistent with our result although we do not observe a downward shift in $T_c$.

We note that, in addition to strangeness neutrality, heavy-ion collisions are also typically characterized by a fixed charge-to-baryon ratio of $n_Q / n_B = 0.4$, reflecting the nucleon content of the colliding nuclei.
This induces a small non-zero $\mu_Q$ in addition to a positive $\mu_S$.
HRG model estimates yield $\mu_Q \approx -0.025\, \mu_B$ at the chemical freeze-out stage~\cite{Vovchenko:2019pjl}.
We estimated the additional effect of a non-zero $\mu_Q$ by performing a calculation at $\mu_Q = -0.025\, \mu_B$ and $\mu_S = 0.15\, \mu_B$ and found it to be virtually negligible.

\subsubsection{Pure strangeness direction}

We also consider separately the pure $\mu_S$ direction, holding $\mu_B = 0$ and $\mu_S = 0$.
In this case, the possible existence of the CP in our scheme is determined entirely by the entropy density $s(T_0)$ and the strangeness susceptibility $\chi_2^S(T_0)$ at vanishing chemical potentials.
We find that the solution to Eqs.~\eqref{eq:CP_4D} exists but located at negative temperatures when using mean values of the input parameters.
A 39\% fraction of the resulting covariance ellipse does extend into positive $T$ plane, indicating the possibility of a CP in the pure-$\mu_S$ direction. It should be noted however, that the extracted values of $\mu_{S,c} \sim 700$ MeV exceed the kaon mass and place the possible CP into the kaon condensation region, where the method may not be reliable.

\subsection{Critical line in the $\mu_B$-$\mu_Q$ plane}

\begin{figure*}[t]
    \centering
    \includegraphics[width=0.4\textwidth]{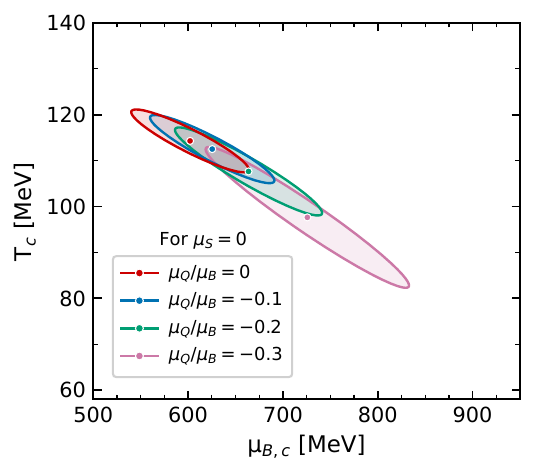}
    \hfill
    \includegraphics[width=0.56\textwidth]{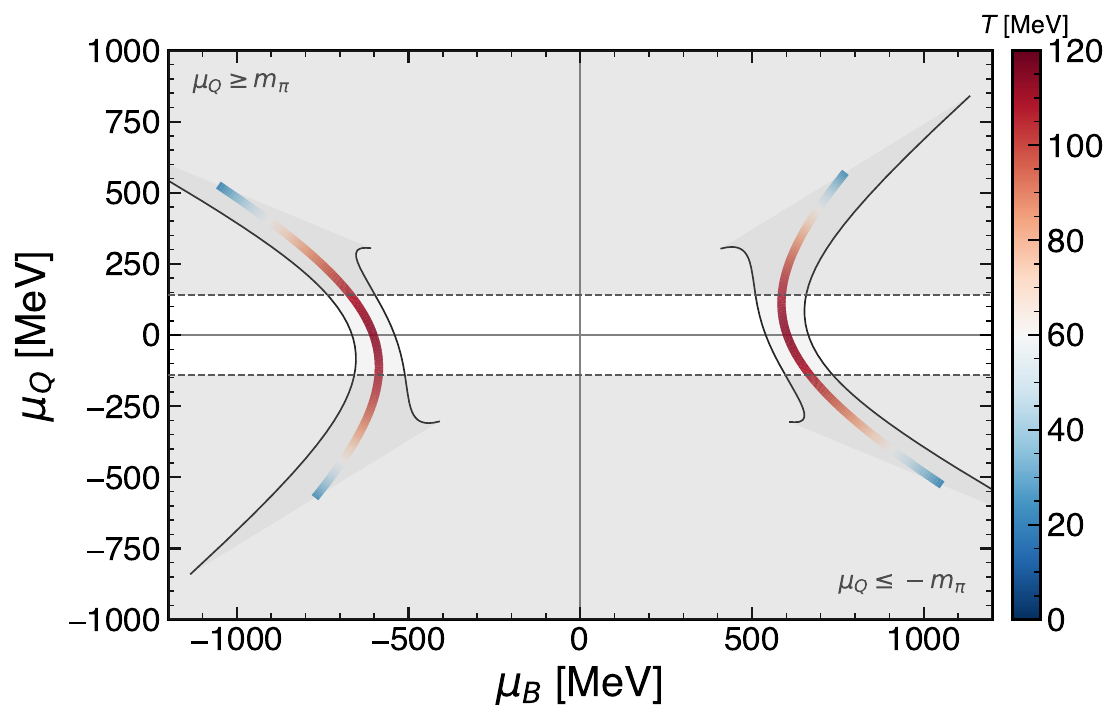}
    \caption{
    Same as Fig.~\ref{fig:muBmuS} but for the $(\mu_B,\mu_Q)$ plane at $\mu_S = 0$.
    The gray shaded region in the right panel corresponds to $|\mu_Q|>m_\pi$, where pion condensation is expected.
    }
    \label{fig:muBmuQ}
\end{figure*}

\subsubsection{$\mu_Q/\mu_B$ scan}

Figure~\ref{fig:muBmuQ} shows the movement of the critical point 
in the $\mu_B-\mu_Q$ plane, which is obtained by fixing $\varphi = 0^\circ$.
The left panel shows the $1\sigma$ uncertainty ellipses at several values of $\mu_Q / \mu_B$, while the the right panel depicts the critical line in $\mu_B$-$\mu_Q$ plane.
We observe only moderate shifts in the CP location when $\mu_Q/\mu_B$ is varied such that the system stays safely outside the pion condensation range, $|\mu_Q| < m_\pi$.
Notably, the critical baryon chemical potential $\mu_{B,c}$ shows an upward shift for negative values of $\mu_Q$ and stays virtually constant for small positive $\mu_Q$ values.
For larger $|\mu_Q|/\mu_B$ ratios, $\mu_{B,c}$ increases regardless of the sign of $\mu_Q$.
This makes the structure of the critical line in the $(\mu_B,\mu_Q)$ plane to appear hyperbolic, in contrast to the elliptical shape in $(\mu_B,\mu_s)$ plane.
The temperature of the critical point decreases with $|\mu_Q| / \mu_B$.
The uncertainties grow as $|\mu_Q| / \mu_B \gtrsim 0.3$.
This reflects the contribution of the input uncertainties from $\chi_{11}^{BQ}$ and $\chi_2^Q$ propagate in addition to those from $s$ and $\chi_2^B$.  
The extracted CP locations for each ratio are given in Table~\ref{tab:muBmuQ_CP}. 

\subsubsection{Charge neutrality and weak equilibrium}

Non-zero charge chemical potential is relevant for astrophysical applications, such as neutron star mergers.
The relevant conditions typically reflect charge neutrality and (approximate) beta equilibrium.
We assume that strangeness is in equilibrium and set $\mu_S = 0$.
The charge chemical potential should be determined from the conditions of charge neutrality, which should include QCD and lepton contributions, $n_Q + n_{\rm leptons} = 0$.
Here we neglect the contribution of the leptons to the equation of state and estimate the value of $\mu_Q$ from the $n_Q = 0$ condition within the HRG model, giving $\mu_Q \approx -(0.05$--$0.1) \,\mu_B$.

We therefore proceed by estimating the CP along the [$\mu_Q \approx -(0.05$--$0.1) \,\mu_B$, $\mu_S = 0$] directions. 
The directions estimated in this way should not be interpreted as exact but rather as representative, motivated by phenomenologically relevant conditions.
Assuming that $\mu_Q = -0.1\, \mu_B$ produces the strongest possible effect of non-zero $\mu_Q$ on the CP location under neutron star merger conditions, we obtain $(T_c, \mu_{B,c}) = (112 \pm 7, 629 \pm 6)$~MeV, representing a small $(\Delta T_c, \Delta \mu_{B,c}) = (-1.8 \pm 0.7, 23 \pm 5)$~MeV shift of the CP location relative to the pure-$\mu_B$ case. 
Our results therefore indicate that, if the CP exists in the pure $\mu_B$ direction at 600, then the CP and the associated first-order phase transition are preserved in the isospin-asymmetric matter relevant for neutron star mergers.

\subsubsection{Large isospin asymmetry and early Universe}

We also explore large $|\mu_Q|/\mu_B$ values corresponding to large isospin asymmetry and moderate baryon densities.
Such a scenario is possible in the early Universe for large lepton flavor asymmetries \cite{Wygas:2018otj,Vovchenko:2020crk,Formaggio:2025nde} which induce non-zero $\mu_B$ and $|\mu_Q| > \mu_B$, and the relevant question is whether cosmic trajectories can cross a first-order phase transition~\cite{Gao:2021nwz,DiClemente:2025awt}.
We find that the CP disappears~($T_c$ becomes negative) for $|\mu_Q| /\mu_B \gtrsim 0.8$ and does not reappear.
In contrast to the pure $\mu_S$ direction, in the pure $\mu_Q$ direction the equations~\eqref{eq:CP_4D} have no solution even for negative $T_c$, at least for the mean values of the parameters.
This is reflected by a hyperbolic structure of the critical curve in the $\mu_B-\mu_Q$ plane, shown in the right panel of Fig.~\ref{fig:muBmuQ}.

Our results here, therefore, are consistent with the absence of a first-order phase transition along the cosmological trajectories in the early Universe.
We note that while our analysis suggests that the pure-$\mu_B$ CP disappears as $|\mu_Q| / \mu_B$ increases, it does not necessarily rule out the existence of a CP at large absolute values of the chemical potentials beyond the reach of the expansion.

\subsection{Critical surface in the $\mu_B$-$\mu_Q$-$\mu_S$ plane}

\begin{table*}[ht]
\centering
\begin{tabular}{|c|c|c|c|c|c|}
\hline
$\mu_Q/\mu_B$ & $T_{0c}$ (MeV) & $T_c$ (MeV) & $\mu_{Bc}$ (MeV) & $\Delta T_c$ (MeV) & $\Delta \mu_{B,c}$ (MeV) \\
\hline
$-0.30$ & $140.7 \pm 2.0$ & $97.6 \pm 15.3$  & $725.7 \pm 107.3$ & $-16.7 \pm 8.9$ & $123.7 \pm 52.8$ \\
$-0.25$ & $140.7 \pm 2.0$ & $103.5 \pm 11.8$ & $690.9 \pm 89.2$  & $-10.8 \pm 5.2$ & $88.8 \pm 33.1$  \\
$-0.20$ & $140.8 \pm 2.0$ & $107.6 \pm 9.6$  & $663.7 \pm 77.6$  & $-6.7 \pm 3.0$  & $61.7 \pm 20.0$  \\
$-0.15$ & $140.8 \pm 2.0$ & $110.5 \pm 8.2$  & $642.3 \pm 70.2$  & $-3.7 \pm 1.6$  & $40.2 \pm 11.2$  \\
$-0.10$ & $140.8 \pm 2.0$ & $112.5 \pm 7.4$  & $625.4 \pm 65.6$  & $-1.8 \pm 0.7$  & $23.4 \pm 5.5$   \\
$-0.05$ & $140.9 \pm 2.0$ & $113.7 \pm 7.0$  & $612.2 \pm 63.0$  & $-0.6 \pm 0.2$  & $10.1 \pm 1.9$   \\
$0$     & $140.9 \pm 1.9$ & $114.3 \pm 6.9$  & $602.1 \pm 62.0$  & $0.0 \pm 0.0$   & $0.0 \pm 0.0$    \\
$0.05$  & $141.0 \pm 1.9$ & $114.3 \pm 7.0$  & $594.6 \pm 62.4$  & $0.0 \pm 0.2$   & $-7.5 \pm 1.1$   \\
$0.10$  & $141.1 \pm 1.9$ & $113.8 \pm 7.3$  & $589.6 \pm 63.8$  & $-0.5 \pm 0.5$  & $-12.5 \pm 2.7$  \\
$0.15$  & $141.1 \pm 1.9$ & $112.8 \pm 7.9$  & $586.7 \pm 66.4$  & $-1.5 \pm 1.0$  & $-15.3 \pm 5.4$  \\
$0.20$  & $141.2 \pm 1.9$ & $111.3 \pm 8.7$  & $586.0 \pm 70.2$  & $-3.0 \pm 1.9$  & $-16.0 \pm 9.4$  \\
$0.25$  & $141.2 \pm 1.8$ & $109.2 \pm 9.9$  & $587.4 \pm 75.4$  & $-5.1 \pm 3.1$  & $-14.6 \pm 15.0$ \\
$0.30$  & $141.3 \pm 1.8$ & $106.6 \pm 11.5$ & $591.0 \pm 82.2$  & $-7.7 \pm 4.7$  & $-11.1 \pm 22.3$ \\
\hline
\end{tabular}
\caption{The location of the critical point up to $1\sigma$ for different ratios of $\mu_Q/\mu_B$. Here, $\Delta T_c = T_c-T_c(\mu_Q{=}0)$ and $\Delta\mu_{B,c} = \mu_{B,c}-\mu_{B,c}(\mu_Q{=}0)$ are the shifts relative to the pure-$\mu_B$ critical point.}
\label{tab:muBmuQ_CP}
\end{table*}

\begin{figure*}
    \centering
    \includegraphics[width=0.99\textwidth]{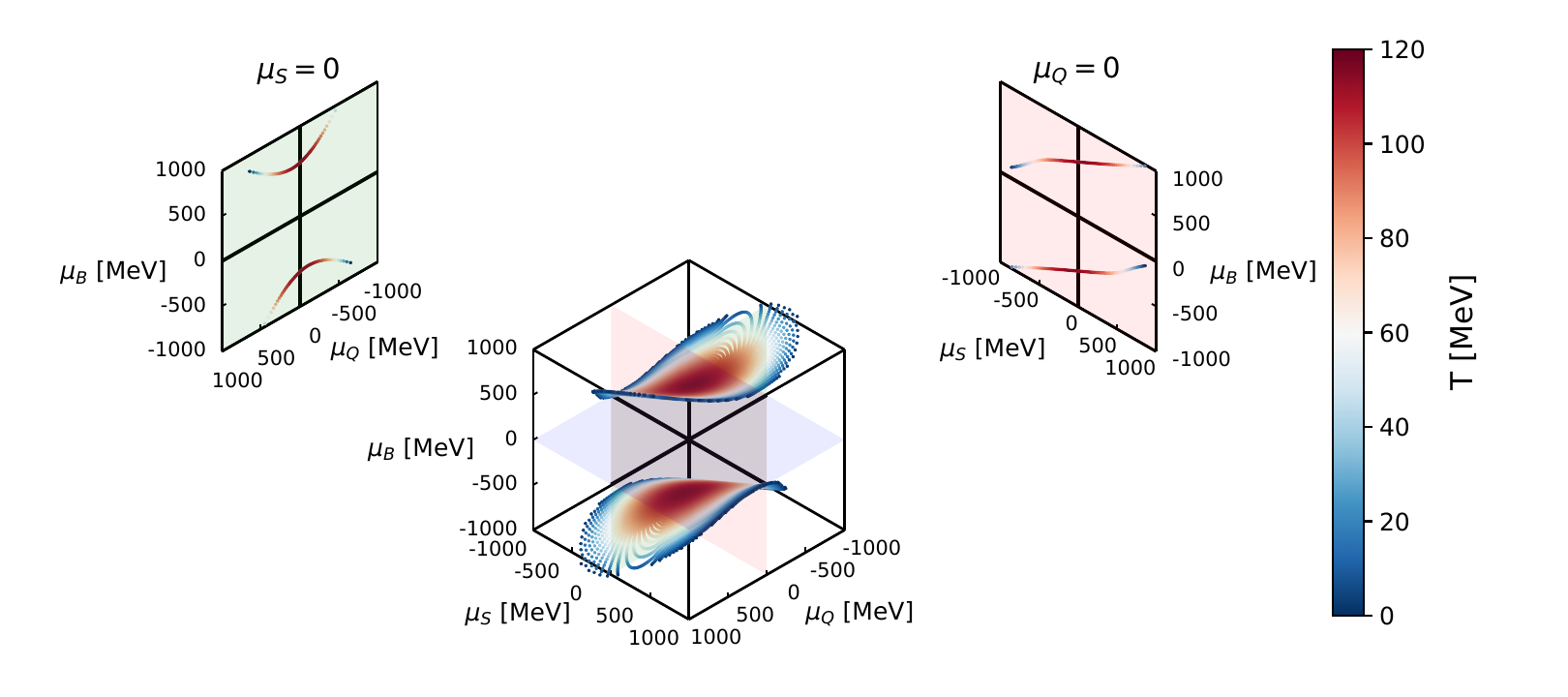}
    \caption{ \justifying The structure of the critical surface in the 4D space of $(T,\mu_B,\mu_Q,\mu_S)$ where the temperature axis is defined based on the color of the surface. The upper-left and upper-right panels show the projections of the critical line in the $\mu_B-\mu_Q$ and $\mu_B-\mu_S$ planes, from Figures~\ref{fig:muBmuS} and~\ref{fig:muBmuQ}, respectively.
    }
    \label{fig:CP_in_4D}
\end{figure*}

We now turn to the critical point structure in the 3D space of chemical potentials.
The critical points form a surface, as follows from the Gibbs phase rule.
Using the mean value of parametrization of the entropy density and conserved charge lattice QCD susceptibilities, in Fig. \ref{fig:CP_in_4D} we show the critical surface in the 3D space of chemical potentials. 
We scan through the angles $\theta$ and $\varphi$ in steps of $5^\circ$ and plot the critical points in the 3D space of chemical potentials. 
The color of the critical points gives the temperature values. 
The charge-conjugation symmetry of the phase diagram is also evident in this plot:
flipping the sign of all the chemical potentials together at any point on the surface results in a different point on the surface. 
Note that we also calculated the uncertainty of the critical point location in the 4D space, but this is not shown in Fig.~\ref{fig:CP_in_4D} to avoid cluttering. 
We verified that the surface structure is preserved within uncertainties. 

To our knowledge, this is the first estimation of the QCD critical  
structure in the 3D space of chemical potentials inferred using lattice QCD results at vanishing chemical potentials.
One limitation of our analysis is that it does not account for the truncation error of the expansion.
The critical surface presented here is thus conditional on the accuracy of the second-order expansion, which corresponds to the scenario where the QCD CP is located at $\mu_{B,c} \sim 600$ MeV in the pure-$\mu_B$ direction.
\section{Conclusions}
\label{sec:Conclusions}
In this work, we extended the constant entropy density contour method to the full three-dimensional space of conserved charge chemical potentials, $(\mu_B,\mu_Q,\mu_S)$. 
This provides, to our knowledge, the first lattice-QCD-based mapping of the QCD critical surface in the full $BQS$ chemical potential space of the (2+1)-flavor QCD. 
Our results are based on the continuum extrapolated entropy density $s(T)$ and second baryon susceptibility $\chi_2^B (T)$ along with the latest continuum extrapolated second-order conserved-charge susceptibilities involving electric charge and strangeness from the Wuppertal--Budapest collaboration \cite{Abuali:2025tbd}. 
To attain the necessary temperature derivatives, we parametrize the input susceptibilities and propagate the lattice uncertainties into the parameter covariance matrix.
By introducing spherical coordinates in the chemical potential space, we reduce the problem to a set of two-dimensional radial expansions, each corresponding to a fixed direction in the $(\mu_B,\mu_Q,\mu_S)$ plane.

We find an approximately hyperbolic critical structure in the $\mu_B-\mu_Q$ plane and an approximately elliptical structure in the $\mu_B-\mu_Q$ plane.
The applicability of our analysis may be bounded in $(\mu_Q,\mu_S)$ directions accordingly by regions where pion or kaon condensation is expected, which our expansion does not describe.

Keeping this caveat in mind, we find no evidence for a critical point at large $\mu_Q$ and small $\mu_B$ and a limited possibility for a critical point at large $\mu_S$ and small $\mu_B$.

The introduction of nonzero $\mu_Q$ and $\mu_S$, and the corresponding shift in the CP location, are relevant to understanding various physical systems, such as heavy-ion collisions and dense astrophysical systems. 
We find that $\mu_{B,c}$ increases by 40-100 MeV along the approximately strangeness neutral direction [$\mu_S \approx (0.15$--$0.33)\, \mu_B$, $\mu_Q \approx 0$] relevant for heavy-ion collisions, while the critical temperature stays essentially unchanged.
In the charge-neutral, weak-equilibrium direction~[$\mu_Q \approx -(0.05$--$0.1) \,\mu_B$, $\mu_S = 0$] relevant for neutron star mergers,
the critical point, and the associated first-order phase transition, remain present at essentially the same location in the $(T,\mu_B)$ plane.
We find no evidence for a critical point at large isospin densities, $|\mu_Q| / \mu_B \gtrsim 1$, relevant for cosmic trajectories in the early Universe.

In the future, we plan to extend the analysis to reconstruct the full 4D equation of state function $P(T,\mu_B,\mu_Q,\mu_S)$ by integrating the entropy density and fixing the integration constant, along the lines of the analysis in the pure-$\mu_B$ direction done in Ref.~\cite{Shah:2026pue}.
The resulting equation of state can then be used in the corresponding simulations of heavy-ion collisions, neutron star mergers, and early Universe evolution.

\section*{Acknowledgments}
This material is based upon work supported by the National Science Foundation under grants No. PHY- 2208724, PHY-2116686 and PHY-2514763, and within the framework of the MUSES collaboration, under Grant
No. OAC-2103680. This material is also based upon work supported by the U.S. Department of Energy, Office of Science, Office of Nuclear Physics, under Award Number DE-SC0022023, as well as by the National Aeronautics and Space Agency (NASA) under Award Number
80NSSC24K0767.
M.H. was supported by the Brazilian logical Development (CNPq) under process No. 313638/2025-0.
V.V. was supported by the U.S. Department of Energy, 
Office of Science, Office of Nuclear Physics, Early Career Research Program under Award Number DE-SC0026065.
%

\bibliography{main}
\begin{appendix}
\widetext 
\setcounter{equation}{0}
\setcounter{figure}{0}
\renewcommand{\theequation}{A.\arabic{equation}}
\renewcommand{\thefigure}{A.\arabic{figure}}
\makeatletter
\section*{Appendix}

\section{Parametrizations of susceptibilities for entropy contour expansion}
\label{apdx:supplemental}
\onecolumngrid

The critical point analysis of Sec.~\ref{sec:results} requires analytic parametrizations of the second-order conserved-charge susceptibilities $\chi_2^B$, $\chi_2^Q$, $\chi_2^S$, and $\chi_{11}^{QS}$ as functions of temperature at $\mu=0$. The remaining two susceptibilities, $\chi_{11}^{BQ}$ and $\chi_{11}^{BS}$, are obtained from these four via the isospin symmetry relations in Sec.~\ref{sec:results}. All four quantities are described by the common parametric form:
\begin{equation}
\chi_2^A(T) = d_0^A \left(\frac{2m_A}{\pi x}\right)^{\!3/2}
\frac{e^{-m_A/x}}{1+(x/d_1^A)^{d_2^A}}
+ d_3^A\,\frac{e^{-(d_5^A)^4/x^4}}{1+(x/d_1^A)^{-d_2^A}},
\label{eq:chi2A_param}
\end{equation}
where $x = T/(200\,\text{MeV})$ and $A \in \{B, Q, S, QS\}$. This functional form was introduced for $\chi_2^B$ in Ref.~\cite{Shah:2024img}; here we apply the same form to the remaining susceptibilities. The low-temperature behavior of each quantity is governed by the lightest hadron carrying the relevant charge, so the mass scale $m_A$ is set to:
\begin{itemize}
    \item $m_B = m_p/(200\,\text{MeV}) \approx 4.69$ (proton mass) for $A = B$;
    \item $m_Q = m_\pi/(200\,\text{MeV}) = 0.70$ (pion mass) for $A = Q$;
    \item $m_S = m_K/(200\,\text{MeV}) = 2.475$ (kaon mass) for $A = S$ and $A = QS$.
\end{itemize}
The fit procedure follows Ref.~\cite{Shah:2024img}: parameters are determined by $\chi^2$ minimization with a correlated lattice covariance matrix of the form 
\begin{align}
    (\Sigma^A)_{ij} = (\sigma^A_i)(\sigma^A_j)\,\Gamma^{|i-j|},
\end{align}
where $\Gamma = 0.84$ accounts for correlations between neighboring temperature points. The covariance matrix of best-fit parameters is taken as the inverse of half the Hessian of $\chi^2$ at the minimum. The entropy density $s(T)$ and $\chi_2^B$ parametrizations, including their parameter values, are taken directly from Ref.~\cite{Shah:2024img}. The parameters for $\chi_2^Q$, $\chi_2^S$, and $\chi_{11}^{QS}$ are obtained here by fitting the continuum estimates of Ref.~\cite{Abuali:2025tbd} and are reported in the tables below.

\paragraph*{\boldmath $\chi_2^Q$:}

The electric-charge susceptibility is fitted using Eq.~\eqref{eq:chi2A_param} with the pion mass $m_Q = 0.70$. The best-fit parameters and their covariance matrix are given in Tables~\ref{tab:cov:chi2Q}.

\begin{table}[ht]
\centering
\begin{tabular}{|c|c|}
\hline
Parameter & Value \\
\hline
$d_0^Q$ & $0.71650$ \\
$d_1^Q$ & $0.75700$ \\
$d_2^Q$ & $6.0154$ \\
$d_3^Q$ & $0.62958$ \\
$d_5^Q$ & $6.688\times10^{-3}$ \\
\hline
\end{tabular}
\hspace{1cm}
\begin{tabular}{|c|c|c|c|c|c|}
\hline
 & $d_0^Q$ & $d_1^Q$ & $d_2^Q$ & $d_3^Q$ & $d_5^Q$ \\
\hline
$d_0^Q$ & $1.615\times10^{-3}$ & $-1.486\times10^{-5}$ & $1.044\times10^{-2}$ & $-3.088\times10^{-4}$ & $5.615\times10^{-4}$ \\
\hline
$d_1^Q$ & $-1.486\times10^{-5}$ & $2.897\times10^{-5}$ & $-7.170\times10^{-4}$ & $4.387\times10^{-5}$ & $-1.071\times10^{-4}$ \\
\hline
$d_2^Q$ & $1.044\times10^{-2}$ & $-7.170\times10^{-4}$ & $9.074\times10^{-2}$ & $-3.050\times10^{-3}$ & $-9.877\times10^{-4}$ \\
\hline
$d_3^Q$ & $-3.088\times10^{-4}$ & $4.387\times10^{-5}$ & $-3.050\times10^{-3}$ & $1.282\times10^{-4}$ & $5.986\times10^{-5}$ \\
\hline
$d_5^Q$ & $5.615\times10^{-4}$ & $-1.071\times10^{-4}$ & $-9.877\times10^{-4}$ & $5.986\times10^{-5}$ & $1.619\times10^{2}$ \\
\hline
\end{tabular}
\caption{\justifying Mean values of parameters describing the lattice QCD data on $\chi_2^Q$ via parametrization~\eqref{eq:chi2A_param}~(left table) and their covariance matrix~(right table)}
\label{tab:cov:chi2Q}
\end{table}

\paragraph*{\boldmath $\chi_2^S$:}

The strangeness susceptibility is fitted using Eq.~\eqref{eq:chi2A_param} with the kaon mass $m_S = 2.475$. The best-fit parameters and their covariance matrix are given in Table~\ref{tab:cov:chi2S}.
\begin{table}[h!]
\centering
\begin{tabular}{|c|c|}
\hline
Parameter & Value \\
\hline
$d_0^S$ & $0.83226$ \\
$d_1^S$ & $0.82958$ \\
$d_2^S$ & $8.0733$ \\
$d_3^S$ & $0.85960$ \\
$d_5^S$ & $0.61863$ \\
\hline
\end{tabular}
\hspace{1cm}
\begin{tabular}{|c|c|c|c|c|c|}
\hline
 & $d_0^S$ & $d_1^S$ & $d_2^S$ & $d_3^S$ & $d_5^S$ \\
\hline
$d_0^S$ & $4.850\times10^{-4}$ & $-1.507\times10^{-4}$ & $-3.318\times10^{-3}$ & $3.154\times10^{-4}$ & $5.365\times10^{-4}$ \\
\hline
$d_1^S$ & $-1.507\times10^{-4}$ & $6.716\times10^{-5}$ & $1.429\times10^{-3}$ & $-1.240\times10^{-4}$ & $-2.208\times10^{-4}$ \\
\hline
$d_2^S$ & $-3.318\times10^{-3}$ & $1.429\times10^{-3}$ & $6.339\times10^{-2}$ & $-4.817\times10^{-3}$ & $-6.195\times10^{-3}$ \\
\hline
$d_3^S$ & $3.154\times10^{-4}$ & $-1.240\times10^{-4}$ & $-4.817\times10^{-3}$ & $3.993\times10^{-4}$ & $5.245\times10^{-4}$ \\
\hline
$d_5^S$ & $5.365\times10^{-4}$ & $-2.208\times10^{-4}$ & $-6.195\times10^{-3}$ & $5.245\times10^{-4}$ & $8.119\times10^{-4}$ \\
\hline
\end{tabular}
\caption{\justifying Mean values of parameters describing the lattice QCD data on $\chi_2^S$ via parametrization~\eqref{eq:chi2A_param}~(left table) and their covariance matrix~(right table).}
\label{tab:cov:chi2S}
\end{table}
\paragraph*{\boldmath $\chi_{11}^{QS}$:}

The charge$-$strangeness correlation susceptibility is fitted using Eq.~\eqref{eq:chi2A_param} with the kaon mass $m_{QS} = 2.475$. For $\chi_{11}^{QS}$, the high-temperature suppression parameter $d_5^{QS}$ is held fixed at $\bar{d}_5^{QS} = 0.41437$ 
Hence, only four parameters are free for this susceptibility. The best-fit parameters and their covariance matrix are given in Table~\ref{tab:cov:chi11QS}.
\begin{table}[h!]
\centering
\begin{tabular}{|c|c|}
\hline
Parameter & Value \\
\hline
$d_0^{QS}$ & $0.33503$ \\
$d_1^{QS}$ & $0.86772$ \\
$d_2^{QS}$ & $7.6043$ \\
$d_3^{QS}$ & $0.28085$ \\
$\bar{d}_5^{QS}$ & $0.41437$ (fixed) \\
\hline
\end{tabular}
\hspace{1cm}
\begin{tabular}{|c|c|c|c|c|}
\hline
 & $d_0^{QS}$ & $d_1^{QS}$ & $d_2^{QS}$ & $d_3^{QS}$ \\
\hline
$d_0^{QS}$ & $9.760\times10^{-5}$ & $2.356\times10^{-7}$ & $8.808\times10^{-4}$ & $-1.273\times10^{-5}$ \\
\hline
$d_1^{QS}$ & $2.356\times10^{-7}$ & $6.876\times10^{-6}$ & $-1.870\times10^{-4}$ & $5.619\times10^{-6}$ \\
\hline
$d_2^{QS}$ & $8.808\times10^{-4}$ & $-1.870\times10^{-4}$ & $1.506\times10^{-2}$ & $-2.773\times10^{-4}$ \\
\hline
$d_3^{QS}$ & $-1.273\times10^{-5}$ & $5.619\times10^{-6}$ & $-2.773\times10^{-4}$ & $6.808\times10^{-6}$ \\
\hline
\end{tabular}
\caption{\justifying Mean values of parameters describing the lattice QCD data on $\chi_{11}^{QS}$ via parametrization~\eqref{eq:chi2A_param}~(left table) and their covariance matrix~(right table).}
\label{tab:cov:chi11QS}
\end{table}

\end{appendix}

\end{document}